\documentclass{article}
\usepackage{spconf,amsmath,graphicx}
\usepackage{enumitem}
\usepackage{amsmath}
\usepackage{adjustbox}
\usepackage{graphicx}
\usepackage{xcolor}
\usepackage{hyperref}
\usepackage{placeins}

\title{Ambisonics Encoding For Arbitrary Microphone Arrays Incorporating Residual Channels For Binaural Reproduction}%

\name{Yhonatan Gayer$^1$, Vladimir Tourbabin$^2$, Zamir Ben-Hur$^2$, Jacob Donley$^2$, Boaz Rafaely$^1$}

\address{$^1$School of Electrical and Computer Engineering, Ben Gurion University of the Negev, Beer-Sheva, Israel. \\$^2$Reality Labs Research at Meta, Redmond, WA, USA.}

\begin{document}
\ninept
\maketitle
\begin{abstract}

In the rapidly evolving fields of virtual and augmented reality, accurate spatial audio capture and reproduction are essential. For these applications, Ambisonics has emerged as a standard format. However, existing methods for encoding Ambisonics signals from arbitrary microphone arrays face challenges, such as errors due to the irregular array configurations and limited spatial resolution resulting from a typically small number of microphones.
To address these limitations and challenges, a mathematical framework for studying Ambisonics encoding is presented, highlighting the importance of incorporating the full steering function, and providing a novel measure for predicting the accuracy of encoding each Ambisonics channel from the steering functions alone. Furthermore, novel residual channels are formulated supplementing the Ambisonics channels. A simulation study for several array configurations demonstrates a reduction in binaural error for this approach. 

\end{abstract}

\section{Introduction}
As virtual reality and augmented reality continue to advance and become more prevalent, there is a growing requirement to capture acoustic environments and produce spatial audio \cite{Review-Paper}. 
Ambisonics \cite{book-Ambisonics} has become a common format for spatial audio rendering, as it enables flexible reproduction of binaural signals tailored to individuals by incorporating their HRTFs \cite{HRTF_measurements}, and adapting to changes in head orientation via simple rotation with the Wigner-D Matrix \cite{Wigner-D}.
Ambisonics signals can be computed from spherical microphone array recordings, by the Plane Wave Decomposition (PWD) \cite{PWD-with-Spherical-conv} technique. However, this approach requires a specially designed microphone array \cite{SH_Processing-book}, \cite{em32} and may not be suitable for mobile or wearable arrays used  for virtual and augmented reality applications \cite{Spherical-Ambisonics}.

With the aim of overcoming these limitations, the methods in \cite{Vlad-paper} and \cite{Parametric-ASM-like-paper} propose the encoding of Ambisonics signals from arbitrary microphone array configurations. The latter employs a preliminary stage of parametric audio encoding, while the former uses a simpler approach, solving linear equations for the Ambisonics coefficients. Despite the potential of these methods, they may face shortcomings. These include errors arising from dependency on scene-related estimated parameters and limitations stemming from the asymmetry of microphone array configurations \cite{DOA-non-symmetric-shape}.

In this study, we revisit the challenge of encoding Ambisonics signals from arbitrary array configurations, originally addressed in \cite{Vlad-paper}, by focusing on signal-independent formulation. The objective of this work is to study the impact of array geometry on the Ambisonics encoding error. Furthermore, the integration of encoded Ambisonics into binaural reproduction is also formulated and investigated. Finally, to bridge the gap imposed by poorly-encoded Ambisonics channels, novel residual channels are introduced for higher-quality binaural reproduction. The paper contributions are outlined below:
\begin{itemize}[left=0em, topsep=0pt, itemsep=-0.3ex]
    \item Introduce and demonstrate a novel measure for encoding quality of an Ambisonics channel based on steering vectors only.
    \item Highlight and demonstrate the importance of incorporating a full steering-vector model.
    \item Introduce and demonstrate the incorporation of residual channels that complement the Ambisonics channels and improve performance in binaural reproduction. 
    \item Demonstrate these concepts using three microphone array configurations, spherical, circular and semicircular.
\end{itemize}
\section{Background}
In this section, we lay the mathematical foundation for array processing, Ambisonics encoding, and binaural reproduction, as used in this paper. We use the spherical coordinate system $(r, \theta, \phi)$ for radius, elevation angle, and azimuth angle, respectively. Equations are presented in the frequency domain with wave number $k = \frac{2\pi}{c}f$, where $c$ is the speed of sound and $f$ stands for frequency.

\subsection{Signal Model and Ambisonics}
Consider an arbitrary array comprising $M$ omni-directional microphones, each positioned at coordinates $(r_i, \theta_i, \phi_i)$, $\forall 1 \le i\le M$, with the set of coordinates referred to as $\Omega_M$. Consider also a set of $Q$ plane waves with directions of arrival (DOA) $(\theta_q, \phi_q) \hspace{2mm} \forall \hspace{2mm} 1 \le q \le Q$, denoted as $\Omega_Q$. The array steering matrix is denoted as $\mathbf{V}(k)$ with dimensions $M \times Q$, where each element $\left[ \mathbf{V}(k) \right]_{i,q}$ corresponds to the frequency response of the $i$-th microphone to a plane wave arriving from the DOA $(\theta_q,\phi_q)$.
The signal measured by the microphone can be expressed as:
\begin{equation}
    \textbf{x}(k) = \mathbf{V}(k)\mathbf{s}(k) + \mathbf{n}(k)
\label{eq:mic model}
\end{equation}
where $\mathbf{x}(k) = [ x_1(k) , ... , x_M(k) ]^T$ is a vector of length $M$, each element $x_i(k)  \hspace{2mm} \forall 1 \le i \le M$ represents the signal captured by the $i$-th microphone. $\mathbf{s}(k) = [ s_1(k) ,\dots, s_Q(k) ]^T$ is the sources signal vector of size $Q$, where each element represents the amplitude of a plane wave at the origin. Finally, $\mathbf{n}(k) = [ n_1(k) , ... , n_M(k) ]^T$ is the microphone noise vector of size $M$, assumed to be independently, identically distributed (i.i.d.) and uncorrelated with $\mathbf{s}(k)$.

The Ambisonics signal due to $\mathbf{s}(k)$ in (\ref{eq:mic model}) and the $Q$ plane waves can be represented as follows, see (2.43) in \cite{SH_Processing-book}:
\begin{equation}
\mathbf{a_{nm}}(k) = \mathbf{Y}_{\mathbf{\Omega}_Q}^H \mathbf{s}(k)
\label{eq:a = Y^H s}
\end{equation}
Here, $\mathbf{Y}_{\mathbf{\Omega}_Q} = [\mathbf{y_{00}},\dots,\mathbf{y_{N_aN_a}}]$ denotes the spherical harmonics (SH) matrix of size $Q \times (N_a+1)^2$, where $\mathbf{y_{nm}} = [Y_{nm}(\theta_1,\phi_1),...,Y_{nm}(\theta_Q,\phi_Q)]^T$, $\forall \hspace{2mm} 0 \leq n \leq N_a$, $-n \leq m \leq n$, denoting a vector of size $Q$ that holds the SH functions of order $n$ and degree $m$, at $(\theta_q,\phi_q)$, see Chapter 1 of \cite{SH_Processing-book}.
Additionally, $\mathbf{a_{nm}}(k) = [a_{00}(k), \ldots, a_{N_aN_a}(k)]^T$ has a size of $(N_a+1)^2$ and holds the Ambisonics signals up to order $N_a$.

Ambisonics signals can be employed for rendering binaural signals \cite{book-Ambisonics}, \cite{Ambisonics2binaurals}, \cite{hrft_Ambisonics}:
\begin{equation}
    p^{l,r}(k) = \mathbf{h_{nm}}^{l,r}(k)^T\mathbf{T}^{N_a} \hspace{0.5mm}\mathbf{a_{nm}}(k)
    \label{eq:p=hnmTanm}
\end{equation}
where $p^{l,r}(k)$ denotes the binaural signal, and the HRTF in the SH domain is $\mathbf{h_{nm}}^{l,r}(k) = [h_{00}^{l,r}(k), \dots, h_{N_hN_h}^{l,r}(k)]^T$ of size $(N_h+1)^2$. For (\ref{eq:p=hnmTanm}) to be applicable we truncate either $\mathbf{h_{nm}}^{l,r}(k)$ or $\mathbf{a_{nm}}(k)$ so \( N_a = N_h \). Matrix $\mathbf{T}^{N_a}$ of size $(N_a + 1)^2 \times (N_a + 1)^2$ holds elements: $0, 1, \text{ and } -1$, which are applied to modify and rearrange the elements of $\mathbf{a_{nm}}(k)$ which holds Ambisonics up to order $N_a$ \cite{book-Ambisonics}, \cite{Vlad-paper}.
The indices $l$ and $r$ represent the binaural signal or the HRTF for each ear, left and right, respectively.

\subsection{Ambisonics Encoding from Arbitrary Arrays}
\label{section: Ambisonics Encoding from Arbitrary Arrays}
Ambisonics is typically encoded from spherical arrays or specifically designed array \cite{book-Ambisonics}, \cite{SH_Processing-book}, \cite{em32}. However, for encoding Ambisonics using arbitrary array configurations as in \cite{Vlad-paper} and \cite{Parametric-ASM-like-paper}, we draw inspiration from binaural signal matching (BSM) \cite{BSM-paper}, which utilizes Tikhonov regularization \cite{Tikhinov}. This approach is based on linear mapping from the microphone signals to the $(n,m)$-th Ambisonic signal:
\begin{equation}
    \hat{a}_{nm}(k) = \mathbf{c_{nm}}(k)^H \mathbf{x}(k)
\label{eq:znm=cnm*xnm}
\end{equation}
$\forall \hspace{1mm} 0\le n\le N_a
, -n \le m \le n$. $\mathbf{c_{nm}}(k)$ is a filter coefficients vector of length $M$, and $\hat{a}_{nm}(k)$ denotes the estimated Ambisonics signal, $a_{nm}(k)$.
This approach entails minimizing the following normalized mean squared error (NMSE) function to compute the optimal coefficients:
\begin{equation}
    \begin{aligned}
        \scalebox{1.5}{\(\varepsilon\)}_{\text{Amb}} = E\left[\left\lVert \hat{a}_{nm}(k) - a_{nm}(k)\right\rVert_2 ^2\right]\bigg/E\left[\left\lVert a_{nm}(k) \right\rVert_2 ^2\right]
    \end{aligned}
    \label{eq:error_nm}
\end{equation}
where, $E\left[.\right]$ represents the expectation operator, and $\left\lVert.\right\rVert_2$ denotes the $l2$ vector norm. Aiming to minimize (\ref{eq:error_nm}) we substituting (\ref{eq:mic model}), (\ref{eq:a = Y^H s}) and (\ref{eq:znm=cnm*xnm}) into (\ref{eq:error_nm}). We assume that the noise $\mathbf{n}(k)$ is white such that $\mathbf{R}_n = E[\mathbf{n}(k)\mathbf{n}(k)^H] = \sigma_n^2 \mathbf{I}$, and is uncorrelated with $\mathbf{s}(k)$, and that $\mathbf{R_s}(k) = \sigma_s^2\mathbf{I}$ which corresponds to qualities of a diffuse sound field composed of Q plane waves, leading to:
\begin{equation}
    \scalebox{1.5}{\(\varepsilon\)}_{\text{Amb}} = \frac{\sigma_s^2 \left\lVert\mathbf{V}(k)^H\mathbf{c_{nm}}(k) - \mathbf{y_{nm}}\right\rVert_2^2 + \sigma_n^2 \left\lVert\mathbf{c_{nm}}(k)\right\rVert_2^2}{\sigma_s^2\left\lVert\mathbf{y_{nm}}\right\rVert_2^2}
\label{eq:Ambisonics analytical error}
\end{equation}
Solving (\ref{eq:Ambisonics analytical error}), leads to:
\begin{equation}
    \mathbf{c_{nm}^{opt}}(k) = 
    \left(\mathbf{V}(k) \mathbf{V}(k)^H + \frac{\sigma_n^2}{\sigma_s^2}\mathbf{I}\right)^{-1}\mathbf{V}(k)
    \mathbf{y_{nm}}
\label{eq:cnm_opt2}
\end{equation}
Eq.~\eqref{eq:cnm_opt2} is applicable when the matrix in brackets is invertible. This is typically the case due to the term \(\frac{\sigma_n^2}{\sigma_s^2}\mathbf{I}\).
To obtain an estimation for $a_{nm}(k)$, we can substitute (\ref{eq:cnm_opt2}) into (\ref{eq:znm=cnm*xnm}), leading to the Ambisonics signal matching (ASM) solution. To comply with sampling considerations, the following condition should be maintained \cite{SH_Processing-book}:
\begin{equation}
    (N_a+1)^2 \le M
    \label{cond:(N+1)^2<M}
\end{equation}
While the fundamental formulation in this section has been presented in \cite{Vlad-paper}, current literature lacks (1) an understanding of the limits of performance and the factor affecting performance, and (2) the effect of these limitations on binaural reproduction. A contribution towards bridging these gaps is presented next. 

\section{Theoretical Analysis of Performance}
\label{section:Theoretical Analysis of Performance}
In this section, we explore the limitations inherent in the signal independent method presented in Se. 2.2, aiming to identify the factors affecting these limitations.

\subsection{The effect of Array Geometry}
\label{sec:The effect of Array Geometry}
While the order of reconstructed Ambisonics has been formulated in (\ref{cond:(N+1)^2<M}), can the entire Ambisonics vector be accurately estimated?
Referring to (\ref{eq:Ambisonics analytical error}), it becomes evident that the reconstruction of $a_{nm}(k)$ is closely tied with the projection of $\mathbf{y_{nm}}$ on the column space of $\mathbf{V}(k)^T$. Hence, the reconstruction becomes error-free when the projection of $\mathbf{y_{nm}}$ over the null space of $\mathbf{V}(k)^T$ denoted $\mathbf{V_0}(k)$, equals zero. Thus we formulate a condition for effective Ambisonics reconstruction:
\begin{equation}
    \begin{aligned}
        \xi_{\text{null}} = 10 \log_{10} \left( \frac{\left\lVert\mathbf{V_0}(k)\mathbf{y_{nm}}\right\rVert_2^2}{\left\lVert\mathbf{y_{nm}}\right\rVert_2^2} \right) \le \textnormal{TH}
    \end{aligned}
    \label{cond:yV0V0y<th}
\end{equation}
where $\mathbf{V_0}(k)$ is constructed through singular value decomposition (SVD) \cite{SVD} of $\mathbf{V}(k)^T$, from the eigenvectors having sufficiently small eigen values. Because $\mathbf{V}(k)^T$ is of size $Q\times M$, assuming $M <Q$, the size of $\mathbf{V_0}(k)$ is lower bounded by $Q \times (Q-M)$. To guarantee effective reconstruction, in this work a threshold of $\text{TH} = -10 dB$ is selected.

The factors affecting (\ref{cond:yV0V0y<th}) primarily relate to microphone placement and the wavelength. These aspects affect the dependency of the microphone steering vectors, consequently impacting $\mathbf{V_0}(k)$. 
Preferred microphone placement involves spacing them adequately apart leading to increased spatial variability. This variability in turn should reduce dependence between the microphone steering vectors, thereby avoiding extending of $\mathbf{V_0}(k)$. This should contribute to improved accuracy in Ambisonics reconstruction according to (\ref{cond:yV0V0y<th}).

As the frequency decreases and wavelength increases, spatial variability typically reduces \cite{SH_Processing-book}. This typically leads to a greater dependence between the microphone steering vectors, thus potentially expanding $\mathbf{V_0}(k)$, leading again to reduced accuracy in the Ambisonics encoding.  

In summary, (\ref{cond:yV0V0y<th}) has been introduced as a measure to support an analysis to identify which channels of the Ambisonics vector can be encoded accurately, a process clearly affected by microphone positioning and frequency.

\subsection{The effect of Steering Function Order}
\label{section:The effect of Steering Function Order}
The method presented in \cite{Vlad-paper} proposed to estimate the Ambisonics signals using quantities defined in the SH domain, without the need for a direction sampling grid. This approach can be useful but only if some conditions are maintained, which were not highlighted in the original paper. The steering matrix can be formulated in the SH domain, as follows:
\begin{equation}
    \mathbf{V}(k) = \mathbf{V_{nm}}(k)^T\mathbf{Y}_{\mathbf{\Omega}_Q}^T
    \label{eq:V=VnmY}
\end{equation}
Here, $\mathbf{V_{nm}}(k) = [\mathbf{v_{00}}(k),\dots,\mathbf{v_{N_vN_v}}(k)]^T$ represents the steering matrix in the SH domain of dimensions $(N_v + 1)^2 \times M$, where $N_v$ stands for the order of the steering function, and $\mathbf{Y}_{\mathbf{\Omega}_Q}$ is defined similarly, as in (\ref{eq:a = Y^H s}), but up to the order of $N_v$. During this section we omit $k$ for brevity. For (\ref{eq:V=VnmY}) to hold, and for accurately computing $\mathbf{V_{nm}}(k)$ from $\mathbf{V}(k)$, two conditions must be satisfied. The first is that $N_v$ is sufficiently high, leading to negligible truncation error in the representation of $\mathbf{V}(k)$ using $\mathbf{V_{nm}}(k)$ \cite{SH_Processing-book}. The second, given this sufficiently high $N_v$, the sampling set $\Omega_Q$ must be chosen to satisfy the aliasing-free sampling condition \cite{SH_Processing-book}, which must also satisfy $Q \ge (N_v+1)^2$. In summary, given these two conditions:
\begin{equation}
\begin{aligned}
    \left[\mathbf{v_{nm}} \right]_i \approx 0, \hspace{2mm} & \forall \hspace{2mm} n>N_v, -n \le m \le n \\
    \text{and} \hspace{1mm} & \forall \hspace{1mm} 1 \le i \le M
\end{aligned}
\label{cond:Vn=0 n>N}
\end{equation}
\begin{equation}
\begin{gathered}
    Q \ge (N_v+1)^2 \text{ for  } \Omega_Q, \\
    \text{an aliasing-free sampling set for order } Nv.
    \label{cond:Omega_Q is aliasing free}
\end{gathered}
\end{equation}
Eq. (\ref{eq:V=VnmY}) can be substituted into (\ref{eq:cnm_opt2}) without error.  
If we further assume negligible noise $(\sigma_n^2 \approx 0)$, to encode Ambisonics up to order $N_a$, the formulation becomes similar to the approach outlined in (10) of \cite{Vlad-paper}, which utilizes the pseudo-inverse \cite{pseudo-inverse}, for the case where $N_v$ satisfies $(N_v+1)^2 \ge M$:
\begin{equation}
    \mathbf{\hat{a}_{nm}}  = \mathbf{T}^{N_v}\mathbf{V_{nm}} ^* \left(\mathbf{V_{nm}} ^T(\mathbf{T}^{N_v})^T\mathbf{T}^{N_v}\mathbf{V_{nm}} ^* \right)^{-1}\mathbf{x} 
    \label{eq:Vldis method}
\end{equation}
where $\mathbf{\hat{a}_{nm}} $ is of size $(N_a+1)^2$ holding the estimated Ambisonics coefficients, the matrix $\mathbf{T}^{N_v}$ is the same as in \eqref{eq:p=hnmTanm}, and $[.]*$ denotes the conjugate element-wise operation.

While (\ref{eq:Vldis method}) may be similar to (\ref{eq:cnm_opt2}), the authors of \cite{Vlad-paper} formulated (\ref{eq:Vldis method}) in such a manner that $N_a=N_v$, i.e. the SH order of the steering vector is the same as the Ambisonics order. As argued above, this selected $N_v$ may lead to truncation error and may not satisfy condition (\ref{cond:Vn=0 n>N}). In summary, this section led to formulating a set of requirements related to $M$, $N_a$, $N_v$ and $Q$, given by Eqs. (\ref{cond:(N+1)^2<M}), (\ref{cond:Vn=0 n>N}), and (\ref{cond:Omega_Q is aliasing free}), which can be used as design guidelines to avoid unnecessary error. 

\subsection{Performance in Binaural Reproduction}
Methods like ASM aim to minimize Ambisonics encoding error (\ref{eq:error_nm}) but are often actually used for binaural reproduction. Alternatively, the BSM method \cite{BSM-paper}, directly minimizes Binaural NMSE and can therefore serve as a performance upper bound for binaural reproduction with ASM. This section aims to explore and formulate the gaps between the two.
The HRTF employed in binaural reproduction can be formulated, as follows:
\begin{equation}
    \mathbf{h}^{l,r} = \mathbf{Y}_{\mathbf{\Omega}_Q}\mathbf{h_{nm}}^{l,r}
    \label{eq:h=Yhnm}
\end{equation}
Here $\mathbf{h}^{l,r}  = [h^{l,r}(\theta_1,\phi_1) ,\dots,h^{l,r}(\theta_Q,\phi_Q)]$ represents the HRTF in the space domain, sampled by the set $\mathbf{\Omega}_Q$. For (\ref{eq:h=Yhnm}) to hold, both conditions (\ref{cond:Omega_Q is aliasing free}) and (\ref{cond:Vn=0 n>N}) must be satisfied with respect to $N_h$ and $\mathbf{h_{nm}}$ instead of $N_v$ and $\mathbf{v_{nm}} $ respectively.
Eq. (\ref{eq:h=Yhnm}) is used next to replace $\mathbf{h}^{l,r} $ in the formulation of binaural reproduction using BSM as in \cite{BSM-paper}, then by using the complex conjugate of the SH as $\mathbf{Y}_{\mathbf{\Omega}_Q}^T = \mathbf{T}^{N_h}\mathbf{Y}_{\mathbf{\Omega}_Q}^H$, we get:
\begin{equation}
    \hat{p}_{\text{BSM}}^{l,r}  = (\mathbf{h_{nm}}^{l,r})^T \mathbf{T}^{N_h} \mathbf{Y}_{\mathbf{\Omega}_Q}^H \mathbf{V}^H \left( \mathbf{V}  \mathbf{V} ^H + \frac{\sigma_n^2}{\sigma_s^2}\mathbf{I} \right)^{-1} \mathbf{x} 
\label{eq:p BSM2}
\end{equation}
here, $\hat{p}_{BSM}^{l,r}$ denotes the estimated binaural signals using BSM. This representation is useful, as it can support the decomposition of the binaural signal into two parts. The first is the binaural reproduction using ASM, and the second is a residual signal bridging the difference between ASM and BSM.
To enable this decomposition, we introduce the following representations:
\begin{equation}
    \mathbf{Y}_{\mathbf{\Omega}_Q} = \left[ \mathbf{Y}_{\mathbf{\Omega}_Q}^{\text{ASM}} | \mathbf{Y}_{\mathbf{\Omega}_Q}^{\text{RES}} \right]
    \label{eq:Y=Y_ASM + Y_RES}
\end{equation}
\begin{equation}
    (\mathbf{h}_{\mathbf{nm}}^{\mathbf{l,r}})^T \mathbf{T}^{N_h} = \left[ (\mathbf{h}_{\mathbf{nm}}^{\mathbf{l,r}}{}^{\text{ASM}})^T \mathbf{T}^{N_a} \middle| (\mathbf{h}_{\mathbf{nm}}^{\mathbf{l,r}}{}^{\text{RES}})^T \mathbf{T}^{\text{RES}} \right]
    \label{eq:hnm = hnmASM + hnmRES}
\end{equation}
Here, \( \mathbf{Y}_{\mathbf{\Omega}_Q}^{ASM} \) is a \( Q \times (N_a+1)^2 \) matrix, containing all SH up to order \( N_a \), satisfying condition \eqref{cond:(N+1)^2<M}, while \( \mathbf{\Omega}_Q \) satisfies \eqref{cond:Omega_Q is aliasing free}. It's counterpart, \( \mathbf{Y}_{\mathbf{\Omega}_Q}^{RES} \) is a \( Q \times ((N_h+1)^2 - (N_a+1)^2) \) matrix, incorporating the remaining SH components that span orders from \( N_a+1 \) to \( N_h \).
Similarly, $(\mathbf{h}_{\mathbf{nm}}^{\mathbf{l,r}}{}^{ASM})^T \mathbf{T}^{N_a}$ is of size $(N_a+1)^2$ representing truncated HRTF up to order $N_a$ with elements that are rearranged or sign inverted. The complementary, \( (\mathbf{h}_{\mathbf{nm}}^{\mathbf{l,r}}{}^{RES})^T \mathbf{T}^{RES} \) has dimensions \( (N_h+1)^2 - (N_a+1)^2 \) and encapsulates the residual HRTF harmonics from orders \( N_a+1 \) to \( N_h \), with elements that are rearranged or sign inverted.

Now, substituting (\ref{eq:Y=Y_ASM + Y_RES}) and (\ref{eq:hnm = hnmASM + hnmRES}) into  (\ref{eq:p BSM2}) leads to:
\begin{equation}
    \hat{p}_{\text{BSM}}^{l,r}  = (\mathbf{h}_{\mathbf{nm}}^{\mathbf{l,r}}{}^{\text{ASM}})^T \mathbf{T}^{N_a} \mathbf{\hat{a}_{nm}}  + (\mathbf{h}_{\mathbf{nm}}^{\mathbf{l,r}}{}^{\text{RES}})^T \mathbf{T}^{\text{RES}} \mathbf{r_{nm}} 
\label{eq:p=ASM+RES}
\end{equation}
where the first components in (\ref{eq:p=ASM+RES}) represents ASM as in (\ref{eq:p=hnmTanm}), and the second component the residual signal, complementing ASM to BSM, where $\mathbf{r_{nm}}$ is a vector containing $(N_h+1)^2-(N_a+1)^2$ residual channels, and is given by:
\begin{equation}
    \mathbf{r_{nm}}  = (\mathbf{Y}_{\mathbf{\Omega}_Q}^\text{{RES}})^H \mathbf{V} ^H  \left( \mathbf{V} \mathbf{V} ^H + \frac{\sigma_n^2}{\sigma_s^2}\mathbf{I} \right)^{-1} \mathbf{x} 
\label{eq:rnm=cnm x}
\end{equation}
The residual signal can serve two purposes. First, it represents the gap between ASM and BSM, and when significant it provides an indication that the gap is significant and binaural reproduction using ASM may be significantly inferior to binaural reproduction using BSM. The second purpose is related to signal encoding. The information in the residual signal can be transmitted together with the Ambisonics signal to enhance reproduction. However, the residual signal may not be a standard Ambisonics signal and so may require special treatment. This part is left as a suggestion for future research.

\section{Simulation Study}

This section presents a simulation study to validate and demonstrate the theoretical foundation presented in the previous section. 

The simulation aims at characterization and comparison of three microphone arrays for Ambisonics encoding, while the acoustic scene is assumed to be composed of  a diffuse sound field, following the derivations in Sec. 3. The three arrays under study all have 4 microphones-one spherical with microphones arranged on a rigid sphere of radius $0.1$m, one circular with microphones uniformly arranged on the equator of the same sphere, and one semicircular, with microphones uniformly positioned on half of the equator of the same sphere. All simulations feature a $20$dB signal to microphone noise ratio and display a frequency range of $0.1 - 8$kHz.

\begin{figure}[t]
  \centering
  \begin{minipage}{\linewidth}
    \centering
    \includegraphics[width=4.87cm]{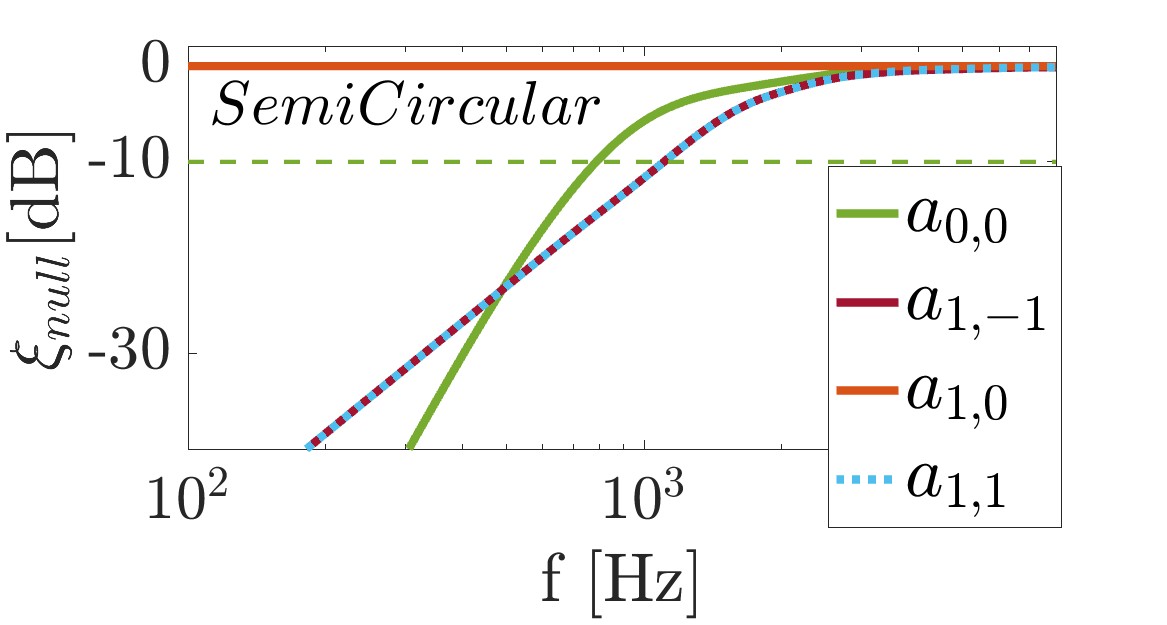}\hspace{-9.5cm}
    \includegraphics[width=4.87cm]{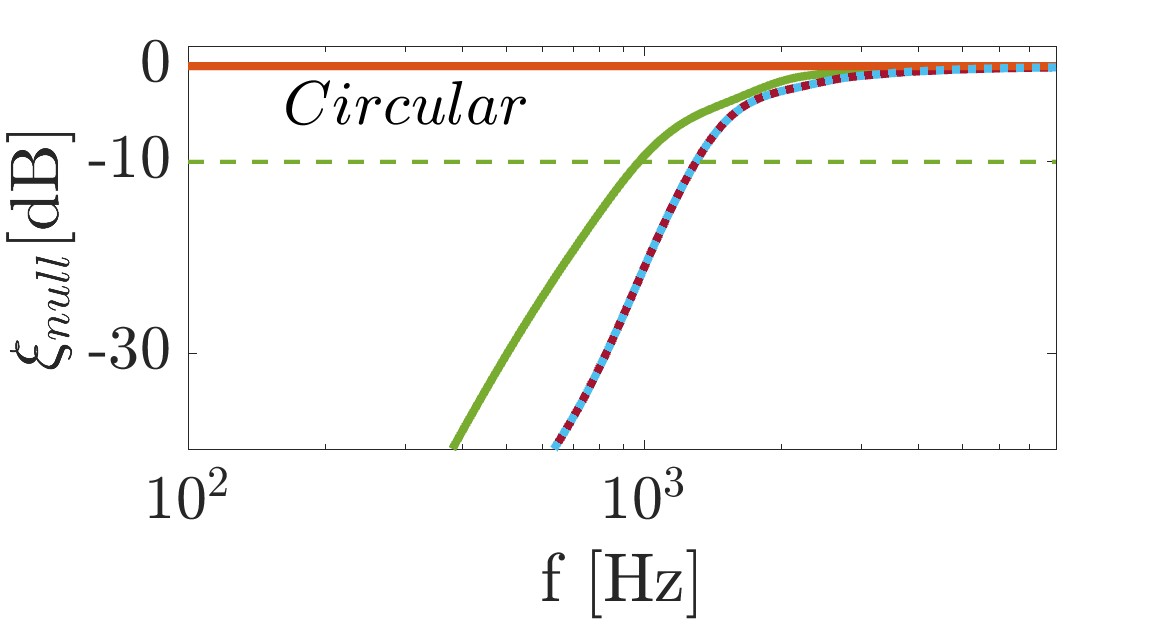}
  \end{minipage}
  \vspace{-0.12cm}
  \begin{minipage}{\linewidth}
    \centering
    \includegraphics[width=0.6\linewidth]{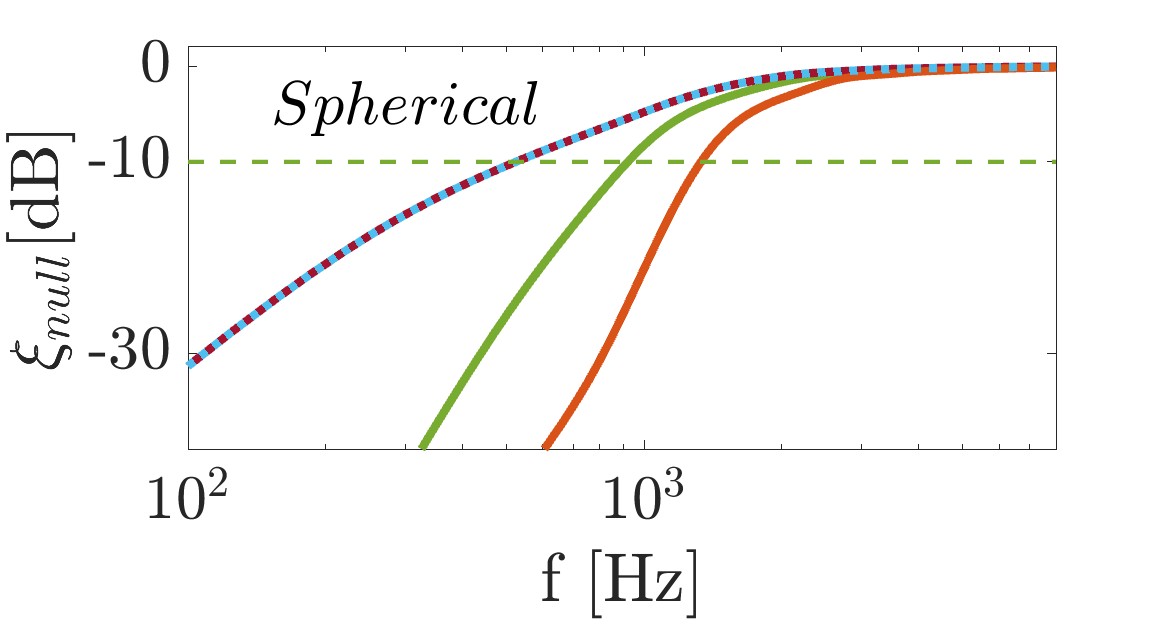}
  \end{minipage}
  
  \caption{\fontsize{9}{11}\selectfont The magnitude of $\xi_{null}$ as defined in (\ref{cond:yV0V0y<th}), as a function of frequency, for $a_{00}(f)$ to $a_{11}(f)$, computed for a $4$-microphone semicircular (top right), circular (top left), and spherical (bottom) arrays.}
  \label{fig:indicator}
\end{figure}

To asses the ability of the arrays to encode all first order Ambisonics signals, the measure $\xi_{null}$ as defined in  (\ref{cond:yV0V0y<th}) is presented in Fig. \ref{fig:indicator}.
The figure shows that the spherical array can facilitate the encoding of all first order Ambisonics signals, as expected, with improved accuracy at the lower frequencies. In contrast, the semicircular and the circular arrays seem unable to reconstruct of \( a_{10}(k) \). In contrast they seem to outperform the spherical array in the reconstruction of $a_{1-1}(k)$ and $a_{11}(k)$. This is attributed to the denser microphone spacing in the horizontal plane \cite{Equatorial-Ambisoncis}. 

\begin{figure}[t]
  \centering
  
  \begin{minipage}{\linewidth}
    \centering
    \includegraphics[width=4.87cm]{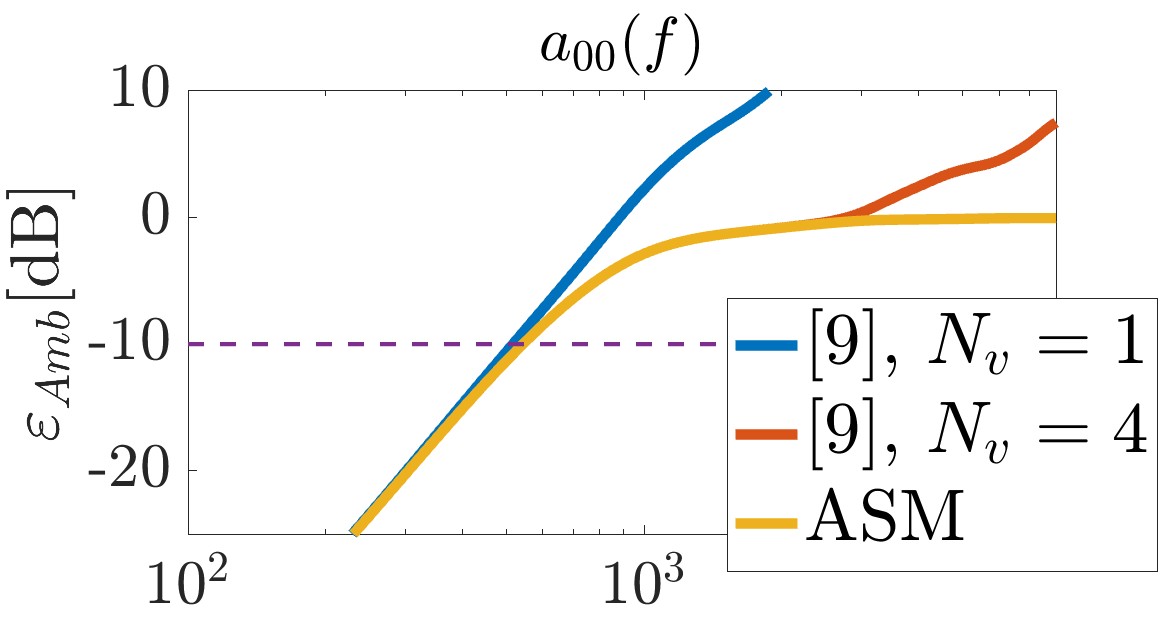}\hspace{-9.5cm}
    \includegraphics[width=4.87cm]{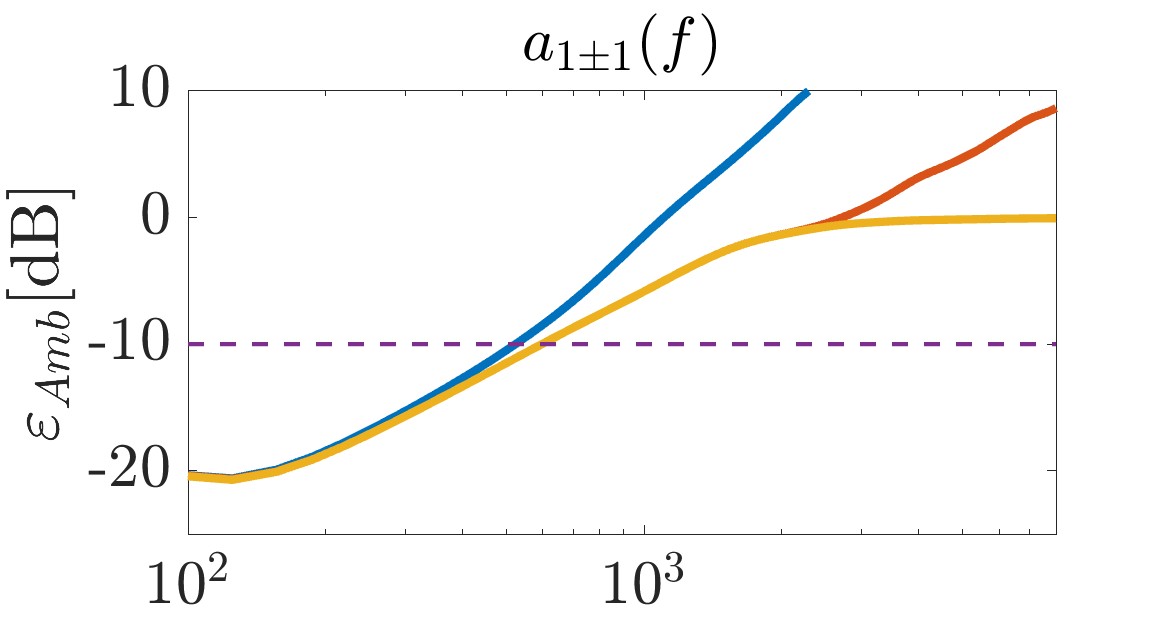}
  \end{minipage}
\vspace{-0.12cm}
  \begin{minipage}{\linewidth}
    \centering
    \includegraphics[width=4.87cm]{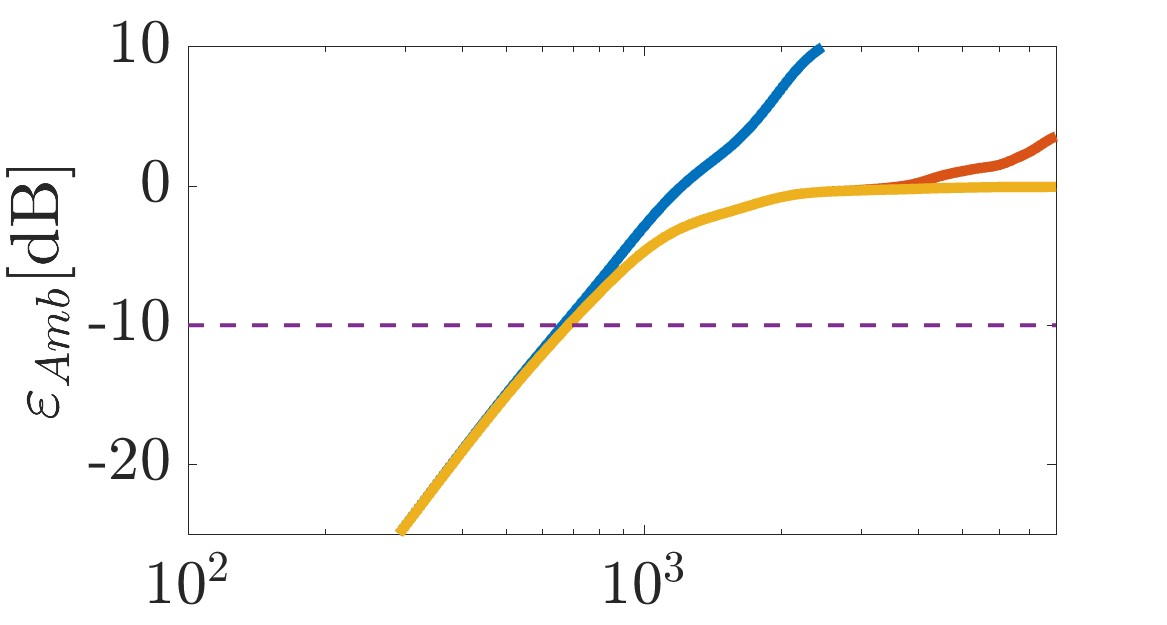}\hspace{-9.5cm}
    \includegraphics[width=4.87cm]{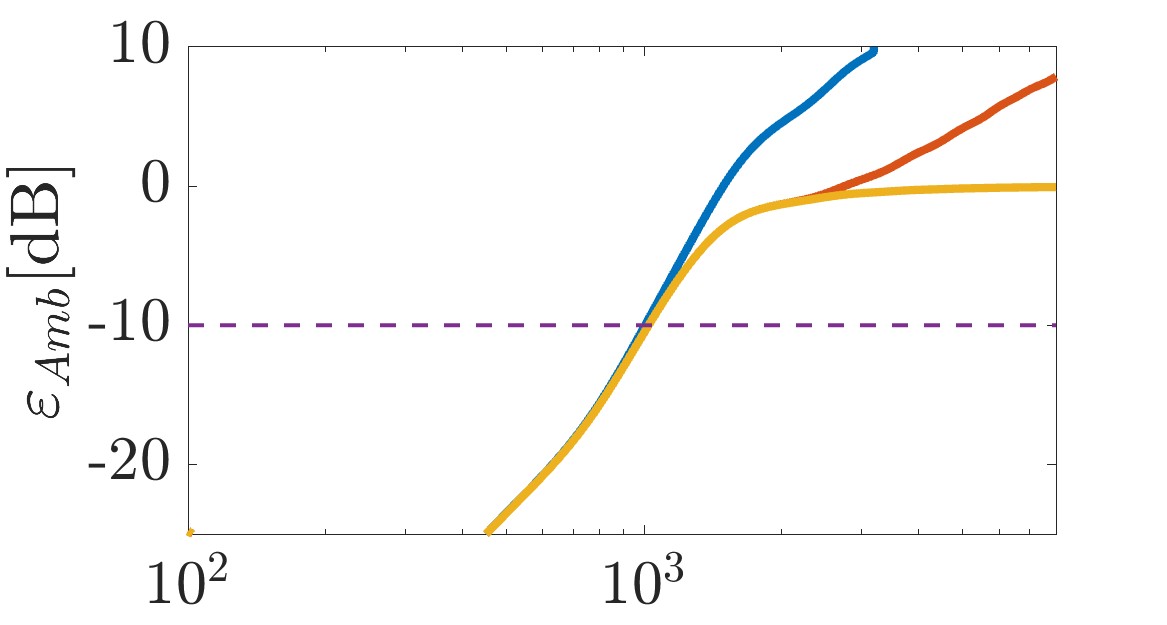}
  \end{minipage}
\vspace{-0.12cm}
  \begin{minipage}{\linewidth}
    \centering
    \includegraphics[width=4.87cm]{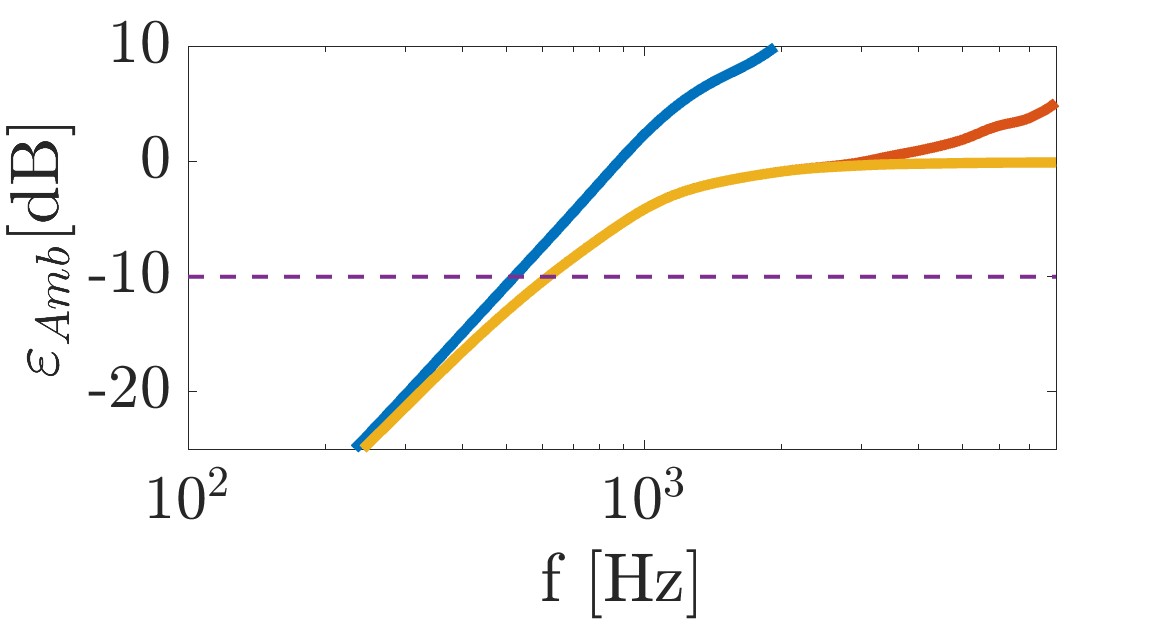}\hspace{-9.5cm}
    \includegraphics[width=4.87cm]{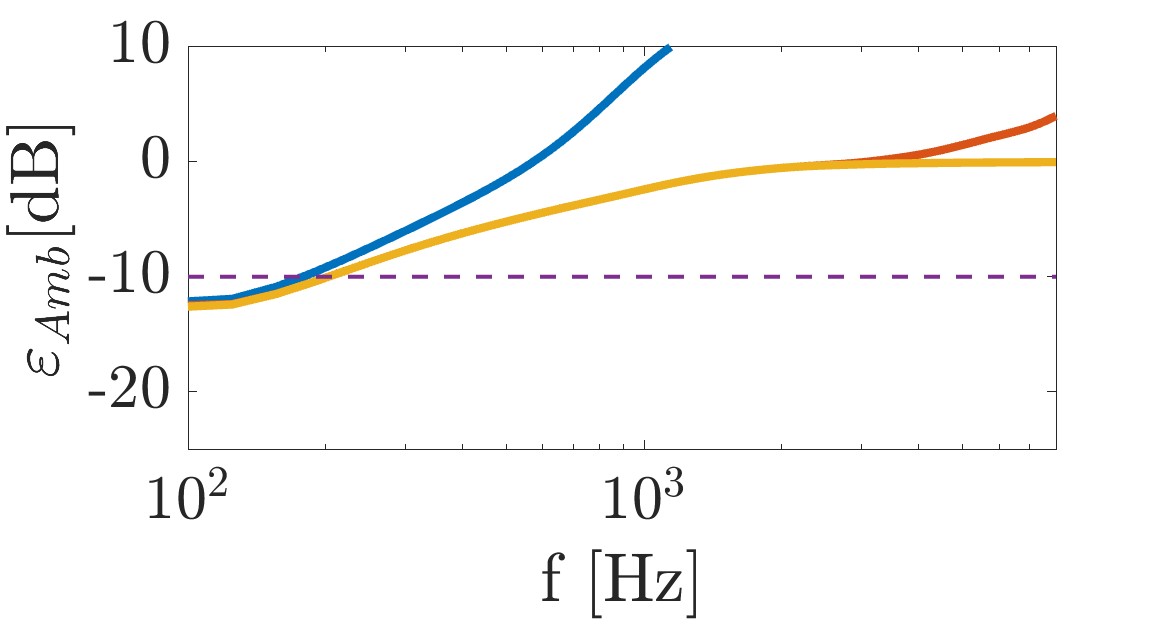}
  \end{minipage}
  
  \caption{ \fontsize{9}{11}\selectfont The error $\scalebox{1.5}{\(\varepsilon\)}_{\text{Amb}}$ in dB as defined in (\ref{eq:Ambisonics analytical error}) using for cases: \(N_v = 1\), \(N_v = 4\), both based on \cite{Vlad-paper}, and ASM. The left plot displays \(a_{1\pm1}(f)\), and the right plot shows \(a_{00}(f)\), both as functions of frequency for a $4$-microphone semicircular (top), circular (middle), and spherical (bottom) arrays.}
  \label{fig:vladimages}
\end{figure}

To explore the impact of steering function order, we examine three cases. The first two cases with a low-order steering function of $N_v = 1$ and $N_v = 4$. For these cases the Ambisonics encoding is performed as in \cite{Vlad-paper} and Eq. (\ref{eq:Vldis method}). In the third case, the complete steering function is used, without any truncation, and Ambisonics encoding is computed using ASM and Eqs. (\ref{eq:cnm_opt2}) and (\ref{eq:znm=cnm*xnm}).
Fig. \ref{fig:vladimages} presents the Ambisonics encoding error, $\varepsilon_{Amb}$ as in (\ref{eq:error_nm}), for the three cases. The figure clearly shows that truncating the steering function to order $N_v = 1$ and $N_v=4$, lead to substantial errors, particularly above $1$kHz. Based on these findings, we advocate for the employment of either a high-order or a complete steering function representation, as emphasized in this study.

In the final part of this simulation study, binaural reproduction is investigated using the binaural NMSE:
\begin{equation}
    \scalebox{1.5}{\(\varepsilon\)}_{\text{bin}} = E\left[\left\lVert {p}^{l,r}(k) - \hat{p}^{l,r}(k)\right\rVert_2 ^2\right]\bigg/E\left[\left\lVert p^{l,r}(k) \right\rVert_2 ^2\right]
\label{eq:error^lr}
\end{equation}
We set \( N_a = N_h = 30\) and use Eqs. \eqref{eq:p=hnmTanm} and \eqref{eq:a = Y^H s} for the reference binaural signal \( p^{l,r}(k) \). Then, three methods are employed for estimating \( \hat{p}^{l,r}(k) \): 1) ASM using $N_a = 1$ incorporating Eqs. \eqref{eq:p=ASM+RES} and \eqref{eq:cnm_opt2} with no residual; 2) ASM but with 5 and 32 residual channels (out of the $(N_h+1)^2-(N_a+1)^2 = 957$ possible residual channels), corresponds to $N_h = 2$ and $N_h = 5$; 3) BSM as an upper performance bound, using \eqref{eq:p BSM2}, which is equivalent to ASM with all residual channels. The Neumann KU100 manikin \cite{HRTF_data_set}, with \( N_h = 30 \) was used for the HRTF. 

\begin{figure}[!ht]
  \centering
    \begin{minipage}{\linewidth}
    \centering
    \includegraphics[width=5cm]{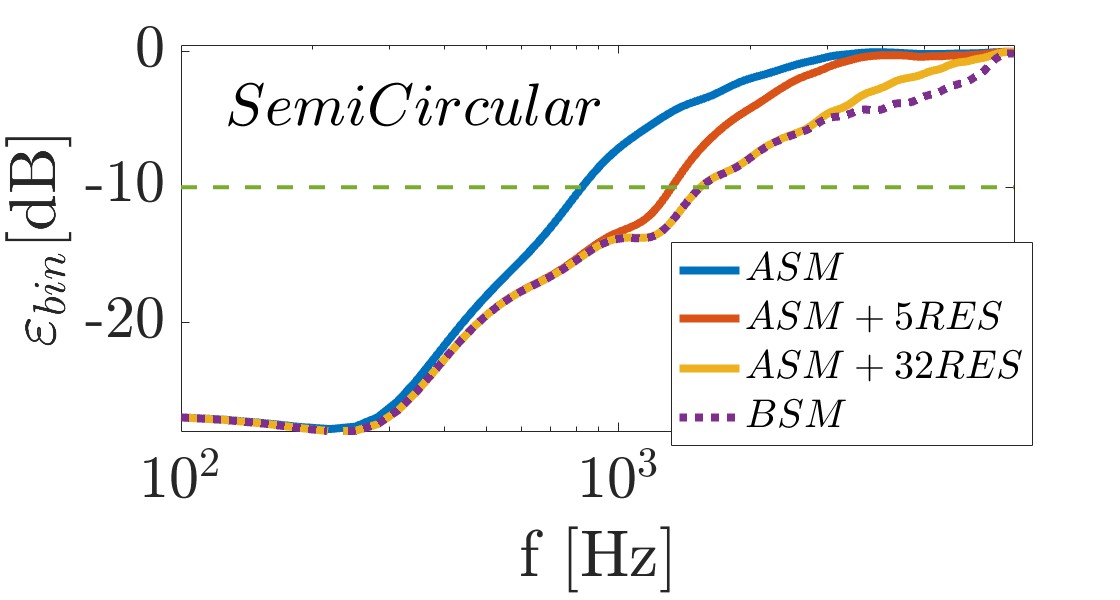}\hspace{-9.7cm}
    \includegraphics[width=5cm]{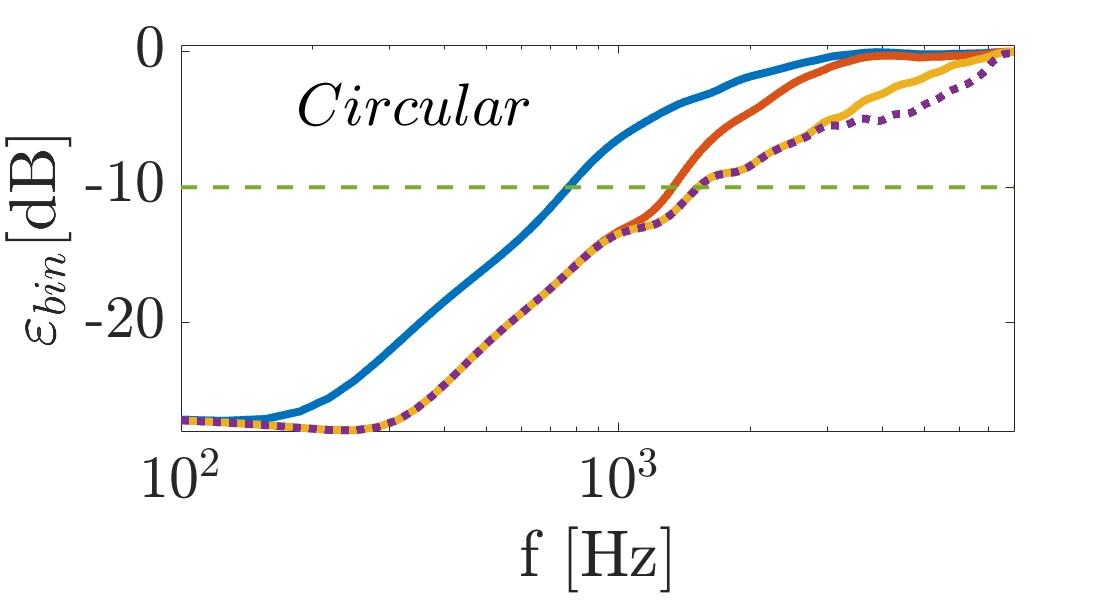}
  \end{minipage}
  \vspace{-0.12cm}
  \begin{minipage}{\linewidth}
    \centering
    \includegraphics[width=0.6\linewidth]{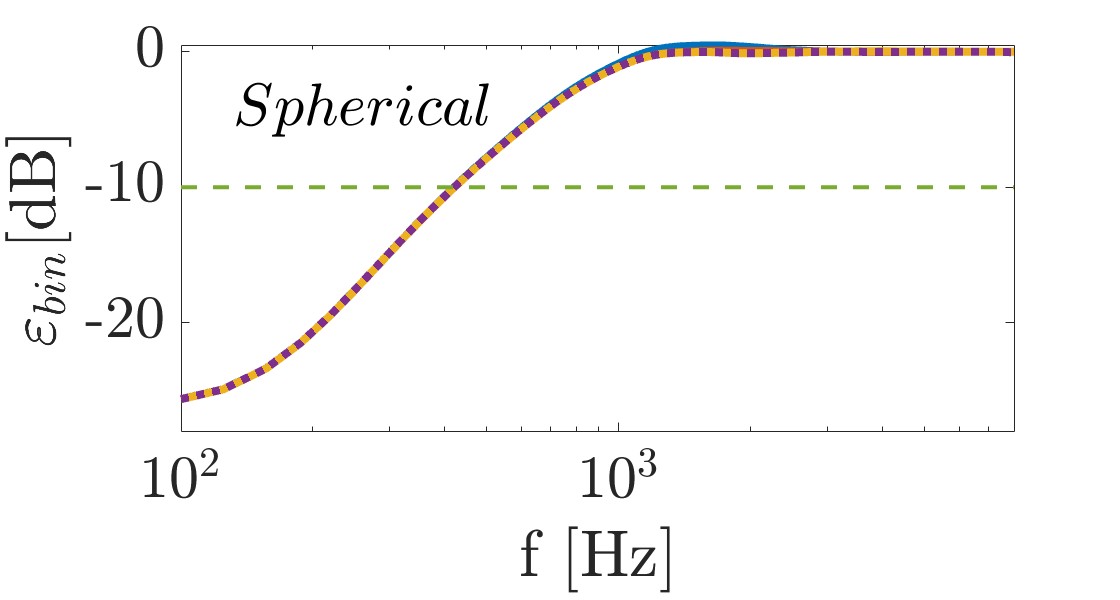}
  \end{minipage}
  \caption{\fontsize{9}{11}\selectfont Binaural error $\scalebox{1.5}{\(\varepsilon\)}_{\text{bin}}$ in dB, as in Eq. (\ref{eq:error^lr}). We examine three methods: 1) ASM. 2) ASM supplemented with $5$ and $32$ residual channels. 3) BSM. All methods are analyzed as functions of frequency for a $4$-microphone semicircular (top left), circular (top right), and spherical (bottom) arrays.}
  \label{fig:Binaural MSE}
\end{figure}

Figure~\ref{fig:Binaural MSE} presents the binaural error, showing that the  incorporation of residual channels into ASM substantially improves performance, for the semicircular and circular arrays. Notably, by employing $32$ residual channels, the performance not only outpaces that of ASM alone but also approximates the optimal performanance of BSM.
As for the spherical array, the close convergence between ASM and BSM is due to the effective full 3D representation of the Ambisonics signals.
The relatively low error in Fig. \ref{fig:Binaural MSE} for the circular and semicircular arrays, despite the error in encoding $a_{10}(k)$, can be explained by the relatively low magnitude of $h_{10}(k)$ in this head positioning. 

\section{Conclusions}
In this study, we explored Ambisonics encoding and binaural reproduction using arbitrary arrays. The study argued theoretically and demonstrated that the limitations in Ambisonics encoding often stem from the shape of the array and the limited representation of the steering matrix.

To overcome error in binaural reproduction within the Ambisonics framework, this paper proposes a novel approach incorporating residual channels transmitted in addition to the Ambisonics signals, emerged as a viable solution for improving binaural rendering. For future work, we propose conducting listening tests and further evaluations to explore the limitations of these residual channels.

\newpage
\bibliographystyle{IEEEbib}
\bibliography{strings,refs}

\end{document}